\documentclass{mn2e}
\usepackage{graphicx}

\title{X-ray pulsar radiation from polar cap heated by back-flow bombardment}
\author[J. Gil, G.I. Melikidze, \& B. Zhang]
       {J.~Gil,$^1$ G.~Melikidze,$^{1,2}$ B.~Zhang$^3$ \\
$^1$Institute of Astronomy, University of Zielona G\'ora, Lubuska 2, 65-265, Zielona G\'ora, Poland\\
$^2$Abastumani Astrophysical Observatory, Al. Kazbegi ave. 2a, 0160, Tbilisi, Georgia\\
$^3$Department of Physics, University of Nevada, Las Vegas, USA}
\date{}

\def\be{\begin{equation}}
\def\ee{\end{equation}}

\def\EB{\hbox{${\rm {\bf E} \times {\bf B}}$}}

\begin{document}

\maketitle

\label{firstpage}

\begin{abstract}
We consider the problem of the thermal X-ray radiation from the hot polar cap of radio pulsars showing evidence of \EB\ subpulse drift in radio band.
In our recent Paper I, using the partially screened gap (PSG) model of inner acceleration region we derived a simple relationship between the drift
rate of subpulses observed in a radio-band and the thermal X-ray luminosity from polar caps heated by the back-flow particle bombardment. This
relationship can be tested for pulsars in which the so-called carousel rotation time $P_4$, reflecting the \EB\ plasma drift, and the thermal X-ray
luminosity $L_x$ from the hot polar cap are known. To test the model we used only two available pulsars: PSRs B0943+10 and B1133+16. They both
satisfied the model prediction, although due to low photon statistics the thermal component could not be firmly identified from the X-ray data.
Nevertheless, these pulsars were at least consistent with PSG pulsar model.

In the present paper we consider two more pulsars: PSRs B0656+14 and B0628-28, whose data have recently become available. In PSR B0656+14 the thermal
radiation from the hot polar cap was clearly detected, and PSR B0628-28 also seems to have such a component.

In all cases for which both $P_4$ and $L_x$ are presently known, the PSG pulsar model seems to be fully confirmed. Other available models of inner
acceleration region fail to explain the observed relationship between radio and X-ray data. The pure vacuum gap model predicts too high $L_x$ and too
low $P_4$, while the space charge limited model predicts too low $L_x$ and the origin of the subpulse drift has no natural explanation.

\end{abstract}

\begin{keywords}
pulsars: pulsars: individual: B0628-28; B0656+14; 0943+10; B1133+16 -- X-rays: thermal
\end{keywords}

\section{Introduction}

Although almost 40 years have passed since the discovery of pulsars, the mechanism of their coherent radio emission is still not known. The theory of
pulsating X-ray emission also demands further development. The puzzling phenomenon of drifting subpulses is widely regarded as a powerful tool for
the investigation of the pulsar radiation mechanism. Recently, this phenomenon received a lot of attention, mostly owing to the newly developed
techniques for the analysis of the pulsar radio emission fluctuations (Edwards \& Stappers 2002, 2003). Using these techniques, Weltevrede, Edwards,
\& Stappers (2006a, WES06 henceforth) presented the results of the systematic, unbiased search for the drifting subpulses and/or phase stationary
intensity modulations in single pulses of a large sample of pulsars. They found that the fraction of pulsars showing evidence of drifting subpulses
is at least 55~\% and concluded that the conditions for the drifting mechanism to work cannot be very different from the emission mechanism of radio
pulsars.

It is therefore likely that the drifting subpulse phenomenon originates from the so-called inner acceleration region right above the polar cap, which
powers the pulsar radiation. In the classical model of Ruderman \& Sutherland (1975; RS75 henceforth) the subpulse-associated spark filaments of
plasma circulate in the pure a Vacuum Gap (VG) around the magnetic axis due to the $\mathbf{E}\times\mathbf{B}$ plasma drift. This model is widely
regarded as a natural and plausible explanation of the drifting subpulse phenomenon, at least qualitatively. On the quantitative level, this model
predicts too high a drifting rate, or too short a period $P_4$ ($\hat P_3$ in the nomenclature introduced by RS75), of the sparks' circulation around
the polar cap, as compared with the observations (e.g. Deshpande \& Rankin, 1999; DR99 henceforth). Also, the predicted heating rate of the polar cap
surface due to the spark-associated back-flow bombardment is too high. The alternative model, namely the space charge limited model (SCLF; e.g. Arons
\& Sharleman 1979), predicts too low a heating rate and has no natural explanation for the phenomenon of drifting subpulses (Zhang \& Harding 2000;
Harding \& Muslimov 2002). However, this model has an advantage over the VG model, namely it is free of the so-called binding energy problem, to
avoid which the VG model requires an ad hoc assumption of the strong, non-dipolar surface magnetic field (for review and more detailed discussion see
Gil \& Melikidze 2002).

Motivated by these observational discrepancies of the otherwise attractive VG model, Gil, Melikidze \& Geppert (2003; GMG03 henceforth) developed
further the idea of the inner acceleration region above the polar cap by including the partial screening caused by the thermionic ions flow from the
surface heated by sparks. We call this kind of the inner acceleration region the "partially screened gap" (PSG henceforth) \footnote{Cheng \&
Ruderman (1980) were the first to consider the PSG model. However, they argued that even with partial screening included, the conditions above the
polar cap are close to pure VG as in RS75. GMG03 demonstrated that the actual thermostatic self-regulation establishes the accelerating potential
drop that may be as low as few percent of that of RS75 value.} Since the PSG potential drop is much lower than that in the RS75 model, the intrinsic
drift rate $P_4$ is compatible with the observations. This is a consequence of the reduced potential drop, partially screened by the thermionic ion
flow from the polar cap surface. In the pure vacuum RS75 gap, the heating of the polar cap is definitely too intense (e.g. Zhang, Harding \& Muslimov
2000; Zhang, Sanwal \& Pavlov 2005; ZSP05 henceforth). On the other hand, the SCLF model predicts too low a heating rate as compared with
observations (Zhang \& Harding 2000; Harding \& Muslimov 2002). Thus, by measuring the thermal X-ray luminosity from heated polar caps one can
potentially reveal the nature of the inner acceleration region in pulsars. This can also help to understand a mechanism of drifting subpulses, which
appears to be a common phenomenon in radio pulsars.

ZSP05 were the first who attempted to test different available models of the inner acceleration region in pulsars, using a concept of the polar cap
heated by the back-flow particle bombardment. They observed the best studied drifting subpulse radio pulsar PSR B0943$+$10 with the {\em XMM-Newton}
observatory and argued that the detected X-ray photons were consistent with PSG formed in the strong, non-dipolar magnetic field just above the
surface of a very small and hot polar cap. Recently Gil, Melikidze \& Zhang (2006 a,b; hereafter Paper I and II, respectively) developed a detailed
model for the thermal X-ray emission from radio drifting pulsars. They applied their model to PSR B0943+10 as well as to PSR B1133+16, which was
observed in X-rays with {\em Chandra} observatory by Kargaltsev, Pavlov \& Garmire (2006, KPG06 henceforth). These authors found that this case is
also consistent with the thermal radiation from a small hot spot, much smaller than the canonical polar cap. PSR B1133$+$16 is almost a twin of PSR
B0943$+$10 in terms of $P$ and $\dot{P}$ values and, interestingly, both pulsars have very similar X-ray signatures, in agreement with the PSG model
(see Table~1 and Fig.~1).

The PSG model can be tested if two observational quantities are known: (i) the circulational period $P_4$ for drifting subpulses observed in the
radio band (also called the pulsar carousel time), and (ii) the X-ray luminosity $L_x$ of thermal black-body (BB) radiation from the hot polar cap
(see Eqs. 2 and 3 below). The above mentioned observations of PSRs B0943+10 and B1133+16 are not decisive. Indeed, due to poor photon statistics,
their spectra can be described by either a thermal model, a non-thermal model, or a combination of the both. In any case, one can pose the upper
limits for the thermal radiation from the hot polar cap from these data, so that the PSG model could be tested at least in the order of magnitude
approximation.

In this paper we include two more pulsars for which values of both $P_4$ and $L_x$ are currently known: PSRs B0656+14 and B0628-28. The former case
was a real breakthrough for our considerations and testing. Indeed, while in the other cases the character of the spectrum was not certain, in this
pulsar (one of the Three Musketeers) the thermal radiation from the hot polar cap was clearly detected (De Luca et al. 2005). PSR B0628-28 was
observed with {\em Chandra} and {\em XMM-Newton} observatories by Tepedelenlio\v{g}lu \& \"Ogelman (2005; hereafter T\"O05). We show that both
pulsars comply the PSG model, increasing the number of pulsars that pass the model test expressed by Eqs.(2) and (3) from two to four. At the moment,
PSRs B0943+10, B1133+16, B0656+14 and B0628-28 are the only pulsars for which both $P_4$ and $L_x$ are known. It is important to show that all of
them follow the theoretical prediction curve in Fig.~1.

\section{PSG model of the inner acceleration region}

The charge depleted inner acceleration region above the polar cap
results from the deviation of a local charge density $\rho$ from the
co-rotational charge density (Goldreich \& Julian 1969) $\rho_{\rm
GJ}=-{\mathbf\Omega}\cdot{\bf B}_s/{2\pi c}\approx{B_s}/{cP}$. For
isolated neutron stars one might expect the surface to consist mainly
of iron formed at the neutron star's birth (e.g. Lai 2001). Therefore,
the charge depletion above the polar cap can result from binding of
the positive $^{56}_{26}$Fe ions (at least partially) in the neutron
star surface. If this is really possible (see Mendin \& Lai 2006, and
Paper II for details), then the positive charges cannot be supplied at
the rate that would compensate the inertial outflow through the light
cylinder. As a result, a significant part of the unipolar potential
drop develops above the polar cap, which can accelerate charged
particles to relativistic energies and power the pulsar radiation
mechanism.

The ignition of cascading production of the electron-positron plasma is crucial for limitation of the growing potential drop across the gap. The
accelerated positrons will leave the acceleration region, while the electrons bombard the polar cap surface, causing a thermal ejection of ions. This
thermal ejection will cause partial screening of the acceleration potential drop $\Delta V$ corresponding to a shielding factor
$\eta=1-\rho_{i}/\rho_{\rm GJ}$ (see GMG03 for details), where $\rho_{i}$ is the charge density of the ejected ions, $\Delta V=\eta({2\pi}/{cP})B_s
h^2$ is the potential drop and $h$ is the height of the acceleration region.  The gap potential drop is completely screened when the total charge
density $\rho=\rho_i+ \rho_+$ reaches the co-rotational value $\rho_{GJ}$.

GMG03 argued that the actual potential drop $\Delta V$ should be thermostatically regulated and there should be established a quasi-equilibrium
state, in which heating due to electron bombardment is balanced by cooling due to thermal radiation. The quasi-equilibrium condition is
$Q_{cool}=Q_{heat}$, where $Q_{cool}=\sigma T_s^4$ is the cooling power surface density by thermal radiation from the polar cap surface and
$Q_{heat}=\gamma m_ec^3n$ is the heating power surface density due to back-flow bombardment, $\gamma=e\Delta V/m_ec^2$ is the Lorentz factor,
$n=n_{GJ}-n_{i}=\eta n_{GJ}$ is the number density of the back-flowing particles that deposit their kinetic energy at the polar cap surface, $\eta$
is the shielding factor, $n_{i}$ is the charge number density of the thermionic ions and $n_{GJ}=\rho_{GJ}/e=1.4\times
10^{11}b\dot{P}_{-15}^{0.5}P^{-0.5}{\rm cm}^{-3}$ is the corotational charge number density. It is straightforward to obtain an expression for the
quasi-equilibrium surface temperature in the form $T_s=(6.2\times 10^4{\rm K})(\dot{P}_{-15}/{P})^{1/4}\eta^{1/2}b^{1/2}h^{1/2}$, where the parameter
$b=B_s/B_d=A_{pc}/A_{bol}$ describes the domination of the local actual surface magnetic field over the canonical dipolar component at the polar cap,
and $\dot{P}_{-15}$ is the normalized period derivative. Here $A_{pc}=\pi r^2_{pc}$ and $A_{bol}=A_p=\pi r^2_p$ is the actual (bolometric) emitting
surface area, with $r_{pc}$ and $r_p$ being the canonical (RS75) and the actual polar cap radius, respectively. Since the typical polar cap
temperature is $T_s \sim 10^6$ K (Paper II), the actual value of $b$ must be much larger than unity, as expected for the highly non-dipolar surface
magnetic fields.

The accelerating potential drop $\Delta V$ and the perpendicular (with respect of the magnetic field lines) electric field $\Delta E$, which causes
\EB\ drift, must be related to each other, and this relationship should be reflected in combined radio and X-ray data of pulsars showing drifting
subpulses. This is basically a conal phenomenon (Rankin 1986), so we can restrict ourselves to the periphery of the polar cap, where these two
potential drops are numerically equal to each other. Moreover, following the original ``pillbox'' method of RS75 we can argue that the tangent
electric field is strong only at the polar cap boundary where $\Delta E=0.5{\Delta V}/{h}=\eta({\pi}/{cP})B_sh$ (see Appendix~A in GMG03 for
details). Due to the \EB\ drift the discharge plasma performs a slow circumferential motion with velocity $v_d=c\Delta E/B_s=\eta\pi h/P$. The time
interval to make one full revolution around the polar cap boundary is $P_4\approx 2\pi r_p/v_d$. One then has
\be
\frac{P_4}{P}=\frac{r_p}{2\eta h}.
\label{P3P}
\ee
If the plasma above the polar cap is fragmented into filaments (sparks), which determine the intensity structure of the instantaneous pulsar radio
beam, then in principle, the circulational periodicity $P_4$ can be measured/estimated from the pattern of the observed drifting subpulses (Deshpande
\& Rankin 1999, Gil \& Sendyk 2003). According to RS75, $P_4=NP_3$, where $N$ is the number of sparks contributing to the drifting subpulse pattern
observed in a given pulsar and $P_3$ is the primary drift periodicity (distance between the observed subpulse drift bands). On the other hand
$N\approx 2\pi r_p/2h=\pi a$, where the complexity parameter can be estimated from the approximate formula $a=5 \dot{P}_{-15}^{0.29}P^{-0.64}$  (Gil
\& Sendyk, 2000; GS00 henceforth). One has to realize that this approximation was derived under a specific assumption concerning the actual surface
magnetic field (see discussion below equation (11) in GS00), and it can give misleading values of $a$ for some untypical pulsars (see discussion in
section 4). However, using this concept we can write the shielding factor in the form $\eta\approx (1/2\pi)(P/P_3)$, which depends only on a
relatively easy-to-measure primary drift periodicity $P_3$. Note also that $P_4/P=a/(2 \eta)$. We show the values of the model parameters obtained
from these equations in Table 1.

\begin{table*}
\begin{minipage}{170mm}
\begin{center}

\footnotesize{ \caption{The observed and the model parameters for the four pulsars.}

\begin{tabular}{l l l l l l l l l l l}

\hline
Name & $a$ & \multicolumn{2}{c}{$P_{4}/P$} & $P_{3}/P$ & \multicolumn{2}{c}{$\eta $} & \multicolumn{2}{c}{$N$} &\multicolumn{2}{c}{$L_{x}\times 10^{-28}\left( erg~s^{-1}\right) $} \\

\hline
PSR B &  & Obs. & Pred. &  & $P/2\pi P_3 $ & $aP/2P_{4}$ &$\left\lfloor P_{4}/P_{3}\right\rfloor $ & $\left\lfloor \pi a\right\rfloor $ & Obs. & Pred. \\

\hline {$0628-28^{\dag}$}& {$7.61$}& {$7_{-1}^{+1}$}& {$6_{-1}^{+1}$}&$0.29_{-0.04}^{+0.04}$ & & {$0.54_{-0.07}^{+0.09}$} &  {$24_{-6}^{+8}$}& {$23$} &  {$287_{-82}^{+152}$} &  {$ 189_{-54}^{+100}$} \\

{$0656+14^{\dag}$} &  {$29.1$} &  {$20^{+1}_{-1}$} & {$18_{-2}^{+2}$} & $0.22^{+0.01}_{-0.01}$ & &  {$0.73^{+0.04}_{-0.03}$} & {$90^{+10}_{-8}$} &  {$91$} &  {$5700_{-561}^{+652}$} &  {$6037_{-561}^{+652}$} \\

{$0943+10$} &  {$6.73$} &  {$37.4_{-1.4}^{+0.4}$} &  {$36_{-6}^{+8}$} & $1.87$ & $ 0.085$ &  {$0.09$} &  {$20_{-1}^{+1}$} &  {$21$} &  {$5.1_{-1.7}^{+0.6}$} &  {$4.7_{-1.3}^{+2.0}$} \\

{$1133+16$} &  {$6.52$} &  {$33_{-3}^{+3}$} &  {$27_{-2}^{+5}$} & $3_{-2}^{+2}$ & $0.05_{-0.02}^{+0.11}$ &  {$ 0.10_{-0.01}^{+0.01}$} &  {$11_{-5}^{+25}$} &  {$20$} &  {$6.8_{-1.3}^{+1.1}$} &  {$5.3_{-0.8}^{+1.1}$} \\

\hline

\multicolumn{10}{l} {Note: $^{\dag}$ As $P_3$ was not measured for these two pulsars we used the estimate of $\eta$ to calculate $P_3/P$.}\\
\end{tabular}}
\end{center}
\end{minipage}
\end{table*}

The X-ray thermal luminosity from the polar cap with a temperature $T_s$ is $L_x=\sigma T_s^4\pi r_p^2=1.2\times 10^{32}(\dot{P}_{-15}/P^3)(\eta
h/r_p)^2$~erg/s, which can be compared with the spin-down power $\dot{E}=I\Omega\dot{\Omega}=3.95 I_{45}\times 10^{31}\dot{P}_{-15}/P^3$~erg/s, where
$I=I_{45}10^{45}$g\ cm$^2$ is the neutron star moment of inertia \footnote{Considering general relativity, the moment of inertia of a neutron star
can be written as $I=0.21 M R^2/(1-2GM/(c^2R))$, where $M$ and $R$ is the neutron star mass and radius, respectively (Ravenhall \& Pethick, 1994).
Taking $M=1.4$ solar masses and $R$ ranging from $8\times 10^5$ to $1.7\times 10^6$ cm, for the softest and stiffest equations of state,
respectively, one obtains the moment of inertia ranging from $7.82\times 10^{44}$ to $2.25\times 10^{45} {\rm g~cm}^2$, respectively.} and
$I_{45}=1^{+1.25}_{-0.22}$. Using equation~(\ref{P3P}) we can derive the formula for thermal X-ray luminosity as
\be
L_x=2.5\times 10^{31}(\dot{P}_{-15}/P^3)(P_4/P)^{-2}
\label{lx},
\ee
or in the simpler
form representing the efficiency with respect to the spin-down
power
\be \frac{L_x}{\dot{E}}=\left(\frac{0.63}{I_{45}}\right)
\left(\frac{P_4}{P}\right)^{-2}
\label{Lx}.
\ee
This equation is very useful for a direct comparison with the observations, since it contains only the observed quantities (although it is subject to
a small uncertainty factor related to the unknown moment of inertia$^2$), and it does not depend on any details of the sparking gap model. It
reflects the fact that both the subpulse drifting rate (due to \EB\ plasma drift) and the polar cap heating rate (due to back-flow bombardment) are
determined by the same physical quantity, which is the potential drop across the inner acceleration region just above the polar cap.

The microscopic properties of PSG model require a more sophisticated analysis, like the one presented in our Paper II. Here we can give simplified
but more intuitive estimate of the screening factor $\eta=(a/2)(P/P_4)=(1/2\pi)/(P/P_3)$ and the number of sparks $N=(P4/P3)=\pi a$, using arguments
based on the complexity parameter $a=r_p/h$ presented in the paragraph below equation (1).

\section{Observational verification}

Table~1 presents the observational data and the predicted values of a number of quantities for four pulsars, which we believe to show clear evidence
of thermal X-ray emission from the spark-heated polar caps as well as they have known values of the circulational subpulse drifting periodicity. The
predicted values of $P_4$ and/or $L_x$ are computed from equation (3). Errors in $L_x$ is taken from the observational papers or derived from the
distance uncertainty (taken from Cordes \& Lazio (2002), except the case of B0656+14 for which it was obtained by Brisken et al. (2003) using the
pulsar parallax), whichever is greater. The relationship expressed by equation (3) is represented by the solid curve in Fig.~1, with two dashed
curves describing the uncertainty in determining of the neutron star moment of inertia$^2$. To save space in Table 1 we give the basic pulsar
parameters ($P, \dot{P}_{-15}, \dot{E}\times 10^{-32}, D$) next to the pulsar name in the paragraphs describing each case below.

{\bf PSR B0943$+$10} ($P=1.099 \ {\rm s}$, $\dot{P}_{-15}=3.49$, $\dot{E} =1.04 \times 10^{32} \ {\rm erg/s}$, $D=0.631_{-0.104}^{+0.113} \ {\rm
kps}$) is the best studied drifting subpulse radio pulsar. As this case, along with PSR B1133+16, was discussed earlier in Papers I and II, we do not
find it necessary to review it again (see Table~1 and Fig.~1). Error bars for $P_4$ were given by Rankin \& Suleymanova (2006), while errors for
$L_x$ were given by ZSP05. See section 4 for discussion.

{\bf PSR B1133$+$16} ($P=1.188 \ {\rm s}$, $\dot{P}_{-15}=3.73$, $\dot{E}=0.88 \times 10^{32} \ {\rm erg/s}$, $D=0.35_{-0.02}^{+0.02} \ {\rm kps}$)
is almost a twin of PSR B0943+10, in both radio and X-ray bands as it was demonstrated in Papers I and II,(see Table~1 and Fig.~1). Error bars for
$P_4$ were given by WES06, while errors for $L_x$ were given by KPG06. See section 4 for discussion.

{\bf PSR B0656$+$14} ($P=0.385 \ {\rm s}$, $\dot{P}_{-15}=55.0$, $\dot{E}=381 \times 10^{32} \ {\rm erg/s}$, $D=0.288^{+0.033}_{-0.027} \ {\rm kps}$)
is one of the famous Three Musketeers, in which the thermal X-ray emission from the hot polar cap was clearly detected (De Luca et al. 2005). This
pulsar is very bright, so the photon statistics are good enough to allow identification of the BB component in the spectrum. As indicated in Table~1,
the X-ray luminosity of this hot-spot BB component is $L_x \sim 5.7 \times 10^{31}$ ergs/s. This value, when inserted into equation (2), returns the
predicted value of $P_4=20.6 P$. Amazingly, Weltevrede et al. (2006b) reported recently the periodicity of $(20\pm 1)P$ associated with the
quasi-periodic amplitude modulation of erratic and strong emission from this pulsar. Thus, it is tempting to interpreted this period as the
circulation time $P_4$. Since there is no doubt about the thermal polar cap emission component, this case greatly strengthens our arguments given for
PSRs B0943+10 and B1133+16, and the equation (3) receives a spectacular confirmation. It is interesting to note that the erratic radio emission
detected by Weltevrede et al.(2006b) is similar to the so-called Q-mode in PSR B0943+10. The low frequency feature in the fluctuation spectra,
identical to the one in the organized B-mode, was found by Rankin \& Suleymanova (2006; see their Fig.~6). Asgekar \& Deshpande (2001;AD01 hereafter)
also detected this feature in the 35-MHz observations of PSR B0943+10 (see their Figs.1 and 2). This simply means that the \EB\ plasma drift is
maintained in both regular (with drifting subpulses observed) and erratic (no drifting subpulses) pulsar emission modes. See some additional
discussion in section 4.

{\bf PSR B0628$-$28} ($P=1.244 \ {\rm s}$, $\dot{P}_{-15}=7.12$, $\dot{E}=1.46 \times 10^{32} \ {\rm erg/s}$, $D=1.444_{-0.277}^{+0.265} \ {\rm
kps}$) is an exceptional pulsar according to T\"O05. Its X-ray luminosity exceeds the maximum efficiency line derived by Possenti et al. (2002) by a
large factor. However, one should note that PSR B0943+10 with its luminosity derived from the PL fit, also exceeds this maximum efficiency (ZSP05,
T\"O05). The BB efficiency $L_x/\dot E \sim 1.9 \times 10^{-2}$ gives the predicted value of $P_4 \sim (6 \pm 1) P$ from equation (2). It is very
interesting that WES06 report the periodicity of $(7 \pm 1) P$ (see Table~1 and Fig.~1). According to the model expressed by equation (3) this
relatively low modulation periodicity (i.e. high modulational frequency) can be interpreted as the circulation time $P_4$. If this is true then PSR
B0628-28 is not an exceptional pulsar at all. It lies on the theoretical curve in Fig.~1 at exactly the right place. This also means that the
observed drift is highly aliased in this pulsar, with $P_3/P$ being considerably lower than 2. As concluded by WES06, this might be the case for most
pulsars. Therefore, all or most features in modulation spectrum frequencies below about 0.2 cycle/P may in fact represent directly the \EB\ plasma
circulation around the pole rather than the apparent subpulse drift periodicity.

\begin{figure}
\includegraphics[scale=0.6]{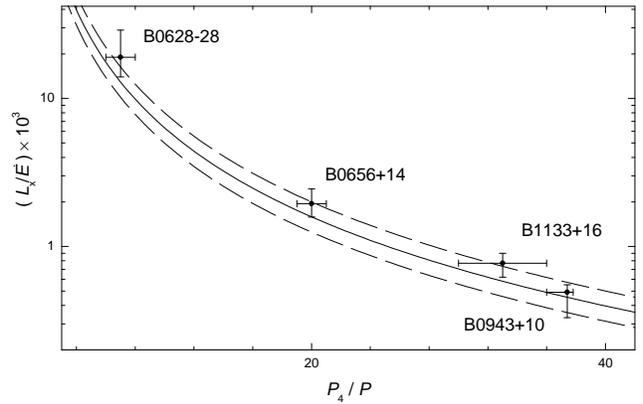}
\caption{The efficiency of thermal X-ray emission from hot polar cap $L_x$ versus circulation period $P_4$ of drifting subpulses in the radio band.
The solid curve represents the prediction of the PSG model (equation 3) with $I_{45}=1$, while the dotted curves correspond to uncertainties in
determining of the moment of inertia$^{1}$. Error bars on $P_4$ and $L_x$ were given by the authors mentioned in the text.}
\end{figure}

\section{Conclusions and Discussion}

Within the partially screened gap (PSG) model of the inner acceleration region in pulsars developed by GMG03, we derived a simple relationship
between the X-ray luminosity $L_x$ from the polar cap heated by sparks and the circulation time $P_4$ of the spark-associated drift detected in radio
band, not necessarily in the form of regularly drifting subpulses. This relationship expresses the fact that both \EB\ drifting rate and polar cap
heating rate are determined by the same value of the available potential drop. In PSRs B0943$+$10, B1133+16, B0628$-$20 and B0654+14, which are the
only pulsars for which both $L_x$ and $P_4$ are known at the moment, the predicted relationship between observational quantities holds very well
(Fig.~1 and Table~1). This suggests that the PSG model may indeed be a reasonable description of the inner accelerator near the polar cap region.
With the abundant radio drifting data (WES06) and the growing number of old pulsars detected in X-rays by {\em XMM-Newton} and {\em Chandra}, the
clean prediction from this model (equation 3) will be unambiguously further tested with more pulsars in the future. PSR B0826-34 with $P_4$ about 14
or 7.5 $P$ (Gupta, Gil, Kijak et al. 2004) and PSR B0834+06 with $P_4$ about 15 $P$, will be examined in the near future.

For the carousel circulation time $P_4$ to be measurable at all, it requires a strong unevenness in the circulating system, maybe a distinguished
group of adjacent sparks or even just a single spark (see also scenario discussed by Gil \& Sendyk (2003). Moreover, it requires this feature to
persist much longer than the circulation time. Such favorable conditions do not occur frequently in pulsars and therefore direct measurements of
$P_4$ are very rare. In principle, in a clean case, using the fluctuation spectra analysis, one should be able to detect the primary feature $P_3$,
reflecting the phase modulation of regularly drifting subpulses, flanked by two symmetrical features corresponding to slower amplitude modulation
associated with carousel circulation. PSR B0943+10 was the first pulsar to show such a model behavior (DR99) and PSR B0834+06 was the second one, as
demonstrated by Asgekar \& Deshpande (2005). The latter authors have also found a direct long period circulational features in both pulsars. For the
B0943+10 they found it in their 35-MHz observations (AD01). In the case of B0834+06 Asgekar \& Deshpande (2005) found an occasional sequence of 64
pulses with much weaker frequency modulation (present in the rest of their data) but with strong long period feature associated with the amplitude
modulation due to the circulation of one or few sparks (see their Fig. 3)\footnote{A close inspection of this sequence of 64 pulses shows also a
presence of even-odd modulation corresponding to the value of $P_3/2$ close to $2$. However, the slope of the secondary drift-bands changes sign,
meaning that $P_3/P$ oscillates around the value of $2$ every $P_4$ periods. Thus, at least in this sequence the subpulse drift in PSR B0834+06 seems
aliased.} Most interestingly, however, Rankin \& Suleymanova (2006) were able to detect a long period circulational feature $P_4$ in the so called
Q-mode erratic emission mode in B0943+10. This apparently first detection of the Q-mode circulation time is very important. Indeed, this fact and
other cases discussed in this paragraph, strongly suggest that no matter the degree of the organization of spark plasma filaments at the polar cap,
the \EB\ drift motion is always performed at the same rate in a given pulsar. The problems is how to reveal this motion.

Different methods of analysis of pulsar intensity fluctuations are sensitive to different effects. The method used WES06 has an obvious advantage of
finding periodicities even in a very weak pulsars, so it resulted in a large increase of pulsars with drifting subpulses and/or periodic intensity
modulation. Generally, WES06 can find only one period and they denote all the periods they find by $P_3$, suggesting that these are primary drift
periodicities. It does not have to be this way at all. In fact, we suggest that at least in three cases their reported values correspond to carousel
circulation times $P_4$. We base our argument mainly on the fact that they satisfy nicely our empirical relationship (Eq. 3 and Fig. 1), without any
obvious selection effect involved. Moreover, in B1133+16 the value of $P_4=(33\pm 3) P$ is close to $(37.4_{-1.4}^{+0.4}) P$ detected in the twin
pulsar B0943+10 (see Paper I for more detailed discussion). In B0656+14 the periodicity of about $20 P$ results from intensity modulation of erratic
spiky emission, similar to the case of Q-mode in B0943+10.

Using a concept of the complexity parameter $a$ (GS00) corresponding to the ratio of the polar cap size to the spark characteristic dimension, we
estimated a number of sparks ,$N$, operating in the inner accelerating regions, as well as values of the screening parameter $\eta$ for the pulsars
discussed in this paper. In the two twin pulsars both $N$ and $\eta$  are almost the same. Is seems trivial since in the approximation we used $a$
depends only on the $P$ and $\dot P$ values, which are close to each other for these two pulsars. However, $N$ can also be found from the ratio of
observed values of $P_4$ and $P_3$, and both estimates are consistent with each other. PSR B0628-28 seems quite similar to the twin pulsars, while in
PSR B0656+14 the number of sparks is 4 times greater, and the screening parameter is quite high (corresponding to about 75 \% of the vacuum potential
drop). This is a result of relatively low $P$ (large polar cap) and unusually high $\dot P$. Thus, either the actual number of sparks is really that
big in this pulsar, or the approximation of GS00 is not good for such a non-typical pulsar. It is not difficult to lower the value of the complexity
parameter and a corresponding number of sparks ($N=\pi a$) by a factor of $2-3$, by considering larger radii of curvature of the actual surface
magnetic field lines, or even the inverse Compton scattering instead of curvature radiation as seed photons for the sparking discharges (see Zhang,
Harding \& Muslimov 2000 and Gil \& Melikidze 2002 for some details).

\section*{Acknowledgments}
We thank Geoff Wright for valuable criticism. We acknowledge the support of the Polish State Committee for scientific research under the grant 1 P03D
029 26. GM was partially supported by Georgian NSF grant ST06/4-096.

{}


\begin{thebibliography}{}
\bibitem{as79} Arons J., Sharleman E.T. 1979, ApJ, 231, 854
\bibitem{aad01} Asgekar A., Deshpande A.A., 2001, MNRAS, 326, 1249
\bibitem{btgg03} Brisken W.F., Thorsett, S.E., Golden A., Goss W.M., 2003, ApJ, 593, L89
\bibitem{cr80} Cheng A.F., Ruderman M.A., 1980, ApJ, 235,576
\bibitem{cl02} Cordes J.M., Lazio T.J.W. 2002, astro-ph/0207156
\bibitem{dl05} De Luca, A., Caraveo, P. A., Mereghetti, S., et al. 2005, ApJ, 623, 1051
\bibitem{dr99} Deshpande A.A., Rankin J.M. 1999, ApJ, 524, 1008 (DR99)
\bibitem{es02} Edwards R.T., Stappers B.W. 2002, A\&A, 393, 733
\bibitem{es03} Edwards R.T., Stappers B.W. 2003, A\&A, 407, 273
\bibitem{gs00} Gil, J., Sendyk M., 2000, ApJ, 541, 351 (GS00)
\bibitem{gm02} Gil J., Melikidze G.I., 2002, ApJ, 577,909
\bibitem{gs03} Gil J., Sendyk M. 2003, ApJ, 585
\bibitem{gmg03} Gil J., Melikidze G.I., Geppert U., 2003, A\&A, 407, 315 (GMG03)
\bibitem{gmz06a} Gil J., Melikidze G., Zhang, B., 2006, A\&A, 457, 5 (Paper I)
\bibitem{gmz06b} Gil J., Melikidze G., Zhang, B., 2006, ApJ, 650, 1048 (Paper II)
\bibitem{gj69}  Goldreich P., Julian H., 1969, ApJ, 157,869
\bibitem{ggks04} Gupta, Y., Gil J., Kijak, J., Sendyk, M., 2004, A\&A, 426, 229
\bibitem{hm02} Harding A., Muslimov A., 2002, ApJ, 568, 862
\bibitem{kpg06} Kargaltsev O., Pavlov G.G., Garmire G.P. 2006, ApJ, 636, 406 (KGP06)
\bibitem{l01} Lai D., 2001, Rev. Mod. Phys., 73, 629
\bibitem{ml06} Medin Z., Lai D., 2006, Phys. Rev. A, 74, 062508
\bibitem{pcc02} Possenti A., Cerutti R., Colpi M., Mereghetti S., 2002, A\&A, 387, 993
\bibitem{r86} Rankin J.M., 1986, ApJ, 301, 901
\bibitem{rjs06} Rankin J.M., Suleymanova S.A., 2006, A\&A, 453, 679
\bibitem{rp94} Ravenhall D.G., Pethick C.J., 1994, ApJ, 424, 846
\bibitem{rs75} Ruderman M.A., Sutherland P.G., 1975, ApJ, 196, 51 (RS75)
\bibitem{to05} Tepedelenlio\v{g}lu E., \"Ogelman H., 2005, ApJ, 630, 57
\bibitem{wes06} Weltevrede P., Edwards R.I., Stappers B.W., 2006a, A\&A, 445, 243 (WES06)
\bibitem{wws06} Weltevrede P., Wright G.A.E., Stappers B.W., Rankin, J. M., 2006b, A\&A, 458, 269
\bibitem{zhm00} Zhang B., Harding A., Muslimov A., 2000, ApJ, 531, L135
\bibitem{zh00} Zhang B., Harding A.K., 2000, ApJ, 532, 1150
\bibitem{zsp05} Zhang B., Sanwal D., Pavlov G.G., 2005, ApJ,624, L109 (ZSP05)
\end{thebibliography}
\end{document}